\documentclass{pasj00}
\begin{document}
\SetRunningHead{M. Arimoto et al.}{HETE-2 Observations of XRF 040916}
\Received{2007/01/09}%{yyyy/mm/dd}
\Accepted{2007/02/28}%{yyyy/mm/dd}

\title{\textit{HETE-2} Observations of the X-Ray Flash XRF 040916}

%%% begin:list of authors
\author{Makoto \textsc{Arimoto}, \altaffilmark{1}
Nobuyuki \textsc{Kawai}, \altaffilmark{1}
Atsumasa \textsc{Yoshida}, \altaffilmark{2}
Toru \textsc{Tamagawa}, \altaffilmark{3} 
}
\author{
Yuji \textsc{Shirasaki}, \altaffilmark{3,4}
Motoko \textsc{Suzuki}, \altaffilmark{3}
Masaru \textsc{Matsuoka}, \altaffilmark{5}
Jun'ichi \textsc{Kotoku}, \altaffilmark{1}
Rie \textsc{Sato}, \altaffilmark{1}
}
\author{
Takashi \textsc{Shimokawabe} \altaffilmark{1}
Nicolas Vasquez \textsc{Pazmino}, \altaffilmark{1}
Takuto \textsc{Ishimura}, \altaffilmark{1}
Yujin \textsc{Nakagawa}, \altaffilmark{2}
}

\author{
Nobuyuki \textsc{Ishikawa}, \altaffilmark{2}
Akina \textsc{Kobayashi}, \altaffilmark{2}
Satoshi \textsc{Sugita}, \altaffilmark{2}
Ichiro \textsc{Takahashi}, \altaffilmark{2}
}
\author{
Makoto \textsc{Kuwahara}, \altaffilmark{3}
Makoto \textsc{Yamauchi}, \altaffilmark{6}
Kunio \textsc{Takagishi}, \altaffilmark{6}
Isamu \textsc{Hatsukade}, \altaffilmark{6}
}
\author{
Jean-Luc \textsc{Atteia}, \altaffilmark{7}
Alexandre \textsc{Pelangeon}, \altaffilmark{7}
Roland \textsc{Vanderspek}, \altaffilmark{8}
Carlo \textsc{Graziani}, \altaffilmark{11}
}
\author{
Gregory \textsc{Prigozhin}, \altaffilmark{8}
Joel \textsc{Villasenor}, \altaffilmark{8}
J. Garrett \textsc{Jernigan}, \altaffilmark{9}
Geoffrey \textsc{B. Crew}, \altaffilmark{8}
}
\author{
Kevin \textsc{Hurley}, \altaffilmark{9}
Takanori \textsc{Sakamoto}, \altaffilmark{12}
George  \textsc{R. Ricker}, \altaffilmark{8}
Stanford \textsc{E. Woosley}, \altaffilmark{16}
}
\author{
Nat \textsc{Butler}, \altaffilmark{8,9}
Al \textsc{Levine} \altaffilmark{8}
John \textsc{P. Doty}, \altaffilmark{8,10}
Timothy \textsc{Q. Donaghy}, \altaffilmark{11}
}

\author{
Donald \textsc{Q. Lamb}, \altaffilmark{11}
Edward \textsc{E. Fenimore}, \altaffilmark{15}
Mark \textsc{Galassi}, \altaffilmark{15}
Michel \textsc{Boer}, \altaffilmark{13}
}
\author{
Jean-Pascal \textsc{Dezalay}, \altaffilmark{13}
Jean-Francios \textsc{Olive}, \altaffilmark{13}
Joao \textsc{Braga}, \altaffilmark{17}
Ravi \textsc{Manchanda} \altaffilmark{18}
}
\author{
and
Graziella \textsc{Pizzichini} \altaffilmark{14}
}

\altaffiltext{1}{Department of Physics, Tokyo Institute of Technology,
2-12-1 Ookayama, \\Meguro-ku, Tokyo 152-8551}
\email{arimoto@hp.phys.titech.ac.jp}
\altaffiltext{2}{Department of Physics and Mathematics, Aoyama Gakuin
University, \\5-10-1 Fuchinobe, Sagamihara, Kanagawa 229-8558}
\altaffiltext{3}{RIKEN, 2-1 Hirosawa, Wako Saitama 351-0198}
\altaffiltext{4}{National Astronomical Observatory of Japan, Osawa,
Mitaka, Tokyo, 181-8588}
\altaffiltext{5}{JAXA, 2-1-1 Sengen, Tsukuba, Ibaraki, 305-8505}
\altaffiltext{6}{Faculty of Engineering, Miyazaki University, Gakuen
Kibanadai Nishi, Miyazaki, 889-2192}
\altaffiltext{7}{LATT, Observatoire Midi-Pyr{\'e}n{\'e}es (CNRS-UPS), 14 Avenue
E. Belin, 31400 Toulouse, France}
\altaffiltext{8}{Center for Space Research, MIT, 77 Vassar Street,
Cambridge, Massachusetts, 02139-4307, USA}
\altaffiltext{9}{Space Sciences Laboratory, University of California ,
Berkeley, California, 94720-7450}
\altaffiltext{10}{Noqsi Aerospace, LTd., 2822 South Nova Road, Pine,
Colorado, 80470, USA}
\altaffiltext{11}{Deparment of Astronomy and Astrophysics, University of
Chicago, \\5640 South Ellis Avenue, Chicago, Illinois 60637, USA}
\altaffiltext{12}{Goddard Space Flight Center, NASA, Greenbelt,
Maryland, 20771, USA}
\altaffiltext{13}{Centre d'Etude Spatiale des Rayonnements,
Observatoire, \\Midi-Pyr{\'e}n{\'e}es, 9 Avenue de Colonel Roche, 31028
Toulouse, France}
\altaffiltext{14}{INAF/IASF Bologna, Via Gobetti 101, 40129 Bologna,
Italy}
\altaffiltext{15}{Los Alamos National Laboratory, P. O. ox 1663, Los
Alamos, NM, 87545, USA}
\altaffiltext{16}{Department of Astronomy and Astrophysics, University
of California at Sata Cruz, \\477 Crark Kerr Hall, Santa Cruz, California,
95064, USA}
\altaffiltext{17}{Instituto Nacional de Pesquisas Espaciais, Avenida Dos
Astronautas 1758, \\ Sa{\~o} Jos{\'e}dos Campos 12227-010, Brazil}
\altaffiltext{18}{Department of Astronomy and Astrophysics, \\Tata
Institute of Fundamental Research,  Homi Bhabha Road, Mumbai, 400-005,
India }

%  \thanks{Example: Present Address is xxxxxxxxxx}}

%\author{Nobuyuki \textsc{Kawai}}
%\affil{Department of Physics, Tokyo Institute of Technology, Meguro-ku,
%  Tokyo 152-8551}\email{nkawai@hp.phys.titech.ac.jp}
%\and
%\author{Atsumasa {\sc Yoshida}}
%\affil{Department of Physics, Aoyama}\email{ccccc@xxx.xxx.xx.xx}
%%% end:list of authors

%%% Please use the following style in case that sorting by 
%%% affilation is impossible. 
%
% \author{%
%   D-Firstname \textsc{D-Familyname}\altaffilmark{1}
%   E-Firstname \textsc{E-Familyname}\altaffilmark{1,2}
%   and
%   F-Firstname \textsc{F-Familyname}\altaffilmark{2}}
% \altaffiltext{1}{Address of Institute}
% \email{ddddd@xxx.xxx.xx.xx}
% \email{eeeee@xxx.xxx.xx.xx}
% \altaffiltext{2}{Address of Institute}

%% `\KeyWords{}' always has to be placed before `\maketitle'.
 %Do NOT move this preamble from here!

\maketitle

\begin{abstract}
A long X-ray flash was detected and localized by the instruments
 aboard the High Energy Transient Explorer II (\textit{HETE-2}) at
 00:03:30 UT on 2004 September 16. The position was reported to the GRB
 Coordinates Network (GCN) approximately 2 hours after the burst. 
This burst consists of two peaks separated by 200 s, with durations of
 110 s and 60 s.
We have analyzed the energy spectra of the 1st and 2nd peaks observed 
with the Wide Field X-Ray Monitor (WXM) and the French Gamma Telescope
 (FREGATE).
We discuss the origin of the 2nd peak in terms of flux
 variabilities and timescales. We find that it is most likely part of the
prompt emission,
and is explained by the long-acting engine model. This  feature
 is similar to some bright X-ray flares detected in the early afterglow
 phase of bursts observed by the Swift satellite.

\end{abstract}

\section{Introduction}
X-ray flashes (XRFs) are generally thought to be a sub-class
of gamma-ray
bursts (GRBs). The main difference between XRFs and GRBs is the
energy of the
emission; the peak energy $E_{\rm peak}$ of XRFs is distributed
in the range from a few keV to  10$-$20 keV
while the one of GRBs is distributed from
 $\sim$20 keV to $\sim$MeV (Barraud et al. 2003). Other properties such as timescales or
features of the
light curve are similar for XRFs and GRBs. Using the
logarithmic fluence ratio
log[$S_{\rm X}$/$S_\gamma$] to categorize bursts, where $S_{\rm X}$ is the 2$-$30
keV fluence 
and $S_\gamma$ is the 30$-$400 keV fluence, Sakamoto et al. (2005)
found that XRFs, X-ray rich GRBs (XRRs), and GRBs form a continuum in
the [$S_{\rm X}$, $S_\gamma$]-plane and in the [$S_{\rm X} / S_\gamma$,
$E_{\rm peak}$]-plane.
This is an evidence that all three kinds of bursts are 
the same phenomenon. Theoretical models which have been proposed
to explain these soft events include off-axis viewing (Yamazaki et
al. 2002), a structured jet (Rossi et al. 2002), and high-z GRBs (Heise et
al. 2001).

In this paper, we report the detection and localization of XRF 040916 by
the $\textit{HETE-2}$ satellite (Ricker et al. 2002) and present the results of a detailed
temporal and spectral analysis.
As this burst has two peaks within a total time interval of
$\sim$350 s, we discuss the origin of the long timescale.

\section{Observation}
\subsection{Localization}
\begin{figure}
  \begin{center}
    \FigureFile(88mm,80mm){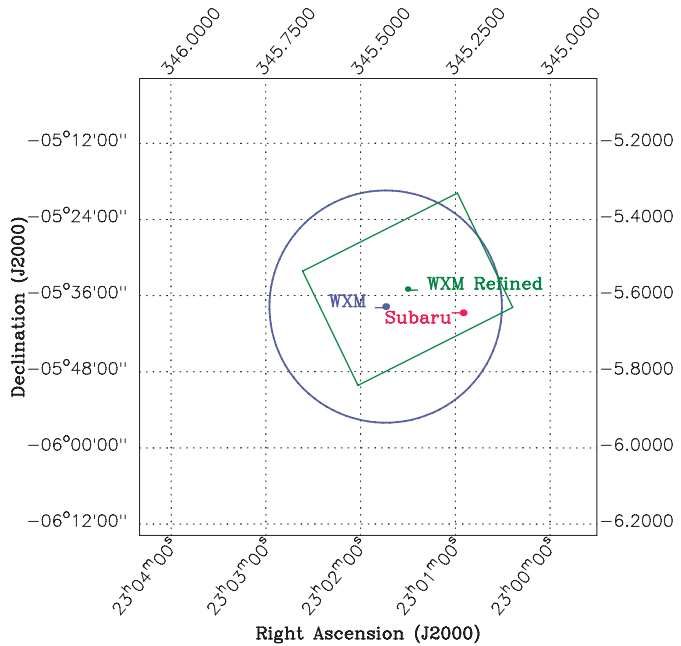}
  \end{center}
  \caption{\textit{HETE-2} WXM localization of XRF 040916. The circle is the initial
 90\% confidence region and the box is the refined 90\%
 confidence region obtained by 
 ground analysis.
 The point labeled ``Subaru'' is the location of the optical afterglow
 (Kosugi et al. 2004, Henden 2004).
 }\label{fig:sample}
\end{figure}

XRF 040916 triggered the  WXM instrument
on 2004 September 16, at 00:03:30 UT (GPS: 779328222.72).
This burst consists of two peaks lasting about 110 s and 60
s separated by a time interval of $\sim$200 s. The \textit{HETE-2} WXM
instrument triggered on the 2nd peak.
The initial burst position was based on a rapid ground
analysis using the WXM data, and was
R.A. = $\timeform{23h01m44s}$, Dec. = $\timeform{-5D37'43''}$ with a
90\% confidence error circle of \timeform{18'} radius. 
All coordinates in this paper are J2000.
This position was reported in a GRB Coordinates Network (GCN) Position
Notice at 02:26:16 UT (Yamazaki et al. 2004) and is shown in Figure \ref{fig:sample}. 
A later, refined ground analysis using the WXM data gave an error
box with the following corners:  (R.A., Dec) =
($\timeform{23h02m01s.68}$, $\timeform{-5D50'09''.6}$),
($\timeform{23h00m23s.76}$, $\timeform{-5D37'51''.6}$),
($\timeform{23h00m58s.80}$, $\timeform{-5D19'51''.6}$),
($\timeform{23h02m36s.72}$, $\timeform{-5D32'09''.6}$)
(90\% confidence region). This was
reported in a GCN Position Notice at 03:58:41 UT(Yamamoto et al. 2004),
and is also shown in Figure \ref{fig:sample}.
This error box is larger than those usually obtained by \textit{HETE-2}
because
XRF 040916 was faint and the Soft X-ray Camera (SXC; 0.5$-$10 keV energy
band; Villasenor et al. 2003) was not operating at the time. No other Interplanetary
Network spacecraft observed this burst, so the localization
could not be refined by triangulation.
Identical localizations were obtained by using the data of the 1st
and 2nd peak separately.

The detection of the optical afterglow was first reported by
Kosugi et al. (2004ab), who found it at R.A. =
$\timeform{23h00m55s.13}$, Dec. =
$\timeform{-5D38'43''.2}$ using SuprimeCam on 
the prime-focus of the Subaru 8.2m telescope (Figure
\ref{fig:sample}).
The afterglow was detected in the z', Ic, Rc, V and B-bands in this observation. Henden et al.
(2004ab) also detected it in
the Ic-band with the NOFS 1.0m telescope. The magnitude of the host galaxy
was estimated to be fainter than Rc=25, which was the magnitude of the afterglow measured 2 days
after the burst.   Despite these observations, no
redshift has been determined for this event.

\subsection{Temporal Properties} \label{sec:temporal}
\begin{figure*}
  \begin{center}
    \FigureFile(120mm,100mm){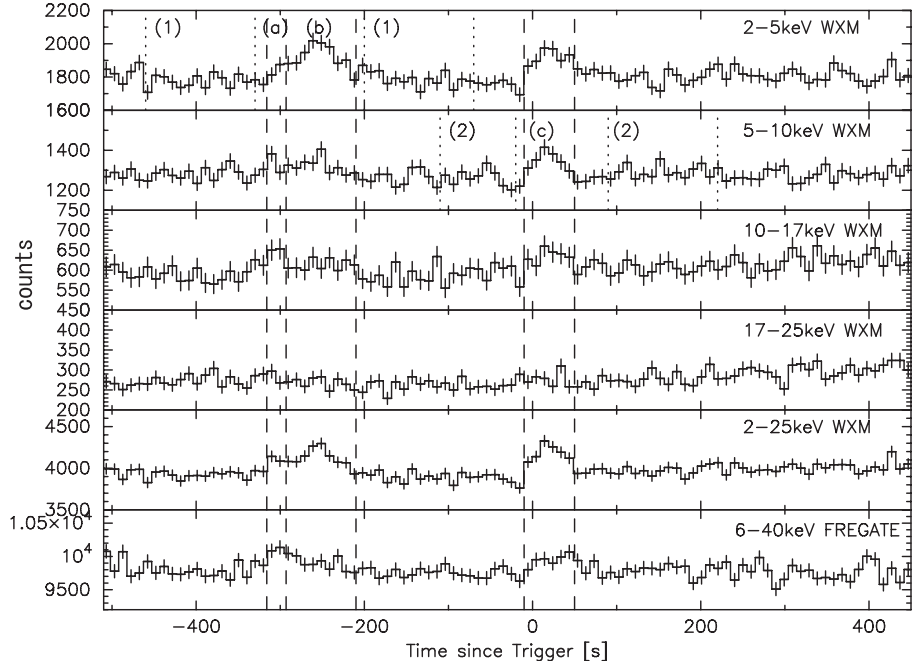}
  \end{center}
  \caption{Light curve of XRF 040916 in five WXM energy bands: 2$-$5,
 5$-$10, 10$-$17, 17$-$25 keV (top to bottom). The light curve has been binned into
 10 second bins. The regions (a) and (b) are the foregrounds of the 1st peak and (c) is
 the foreground of the 2nd peak. Region (1) represents the
 background used for the 1st
 peak and region (2) is the background for the 2nd peak.}\label{fig:thx}
  \label{thx}
\end{figure*}

Figure \ref{fig:thx} shows the light curve of XRF 040916 in five WXM energy
bands (2$-$5, 5$-$10, 10$-$17, 17$-$25, and 2$-$25 keV). There are two peaks in the WXM
bands and no significant emission above 17 keV. Table
\ref{tab:temporal_tot}  gives the 
T$_{50}$ and T$_{90}$ durations in the 2$-$5, 5$-$10, 10$-$25,
and 2$-$25 keV energy bands for the first peak, the second peak, and the entire burst. 

The 1st peak is composed of  two parts :  a hard emission region and a soft
one, referred to as regions (a) and (b), respectively, in figure
\ref{fig:thx}.
This 'hard-to-soft' evolution is typical of GRBs.
The duration of the 2nd peak
tends to be shorter at higher energies,
which is a common feature observed in many GRBs (
Link et al. 1993; Fenimore et al. 1995). 
The duration of the 1st peak in the 10$-$17 keV band is longer than the
one in 2$-$5 keV band because, in the high energy band the 1st peak
consists of two pulses. In the region (a) there are few photons in the 
 2$-$5 keV band while there are certain photon contributions in the
 5$-$10 keV and 10$-$17 keV bands.

\begin{table*}
  \caption{Temporal properties of XRF 040916 (Quoted errors correspond to 1$\sigma$)}\label{tab:temporal_tot}
  \begin{center}
    \begin{tabular}{c|cc|cc|cc}
     \hline
     Region & \multicolumn{2}{|c|}{1st \& 2nd peak} &
     \multicolumn{2}{|c|}{1st peak} &
     \multicolumn{2}{|c}{2nd peak} \\ \hline
     Energy & t$_{90}$ & t$_{50}$ & t$_{90}$ & t$_{50}$ & t$_{90}$ & t$_{50}$ \\ 
	(keV) & (s) & (s) & (s) & (s) & (s) & (s) \\ \hline	
     WXM & & & & & & \\
     2$-$5 & 374.8 $\pm$ 6.2 & 287.5 $\pm$ 3.7 & 103.3 $\pm$ 6.4 & 39.3
     $\pm$ 3.2 & 61.4 $\pm$ 4.1 & 28.3 $\pm$ 1.3\\
     5$-$10 & 350.2 $\pm$ 6.4 & 292.4 $\pm$ 4.1 & 90.9 $\pm$ 1.7 & 50.4 $\pm$ 3.7 & 39.4 $\pm$ 3.4 & 19.7 $\pm$ 2.1  \\
     10$-$17 & 392.0 $\pm$ 17.3 & 325.6$\pm$ 5.2 & 120.5 $\pm$ 5.2 & 67.6
     $\pm$ 4.6 & 44.3 $\pm$ 8.6 & 18.4 $\pm$ 7.2 \\ \hline
     FREGATE & & & & & &\\
     6$-$40 & 349.0 $\pm$ 6.2 & 109.4 $\pm$ 5.4 & 93.4 $\pm$ 13.1 & 
     43.0 $\pm$ 10.3 & 55.3 $\pm$ 7.6 &  30.7 $\pm$ 9.4 \\ \hline
    \end{tabular}
  \end{center}
\end{table*}

\subsection{Emission between the 1st and 2nd peaks}

We calculated the count rates before and after the 1st peak, 
and compared them to the count rate after
the 2nd peak. 
The results show that there is no significant difference between
them.
We also tried to localize the burst using the photons in
the time region between the 1st and 2nd peaks. 
If there is emission from the burst between the peaks,
the resulting localization should be the same as that of the
1st or 2nd peak alone.  However, a significant localization could not
be obtained,
and we conclude that there is no significant emission between the peaks. 
The corresponding 2$\sigma$ upper limit is  1.8 $\times$ 10$^{-9}$ ergs
cm$^{-2}$ s$^{-1}$ (2$-$400 keV).

\subsection{Spectrum} \label{sec:spectrum}

We performed spectral analyses of the two peaks
separately, and also of their sum.
The background regions used were -460 $\leq$ t $\leq$ -330 s and -200
$\leq$ t $\leq$ -70 s for the 1st peak, and -110 $\leq$ t $\leq$ -20
s and 90 $\leq$ t $\leq$ 220 s for the 2nd peak, where t is the
time since trigger.

Two types of data sets ($\textit{burst data}$ and $\textit{survey data}$) are provided
by the  \textit{HETE-2} WXM and FREGATE detectors (Atteia et
al. 2003). The burst data are only
available when the detector triggers on a burst, while the survey
data 
are recorded whenever the \textit{HETE-2} satellite is operating.
The trigger occurred for the second peak of XRF 040916, and consequently both sets of  
data were available for it, whereas only the survey data
were obtained for the 1st peak.

The burst data include time-tagged photon data,
while the survey data produce time-integrated
(4.92 s) data for each wire in the WXM proportional counters.
(The WXM instrument is composed of twelve 1-D position-sensitive
proportional counters.) 
In order to improve the signal to noise ratio, and consequently the
spectral analysis, 
we applied a cut to the WXM photon time- and energy-tagged data (TAG
data),
using only the photons from the pixels on the six
wires of the X-detector and the two wires of the Y-detector that were
illuminated by the burst.
Moreover, as the gain is not uniform at the end of the wires (Shirasaki
et al. 2003), we used only the photon counts that registered in the
center $\pm$ 50 mm region of the wires. 
These \textit{optimized TAG data} were extracted and used for the 2nd peak of XRF 040916.

We used the $\textit{XSPEC v.11.3.0}$ software package 
(Arnaud, K. 1996) to perform the  spectral analysis.
We simultaneously fit the WXM and FREGATE data with the four
following functions; (1) blackbody function, (2)  power-law function,
(3) cutoff power-law function, and (4) Band function (Band et al. 1993).
Table \ref{tab:Spec_fit} shows the results of the spectral analysis of each
peak. 

Although the spectrum is soft, several facts argue against a type I
X-ray burst (XRB) 
as the origin. First, the XRB
emission mechanism is blackbody radiation, and we obtain a
large $\chi^2$ except for the 1st peak (a) when we use a blackbody
model to fit the data. 
Second, XRBs tend to be found in the Galactic plane or globular clusters, and they emit
persistent X-rays; the Galactic latitude of XRF 040916 is
$\textit{b}$ = $\timeform{-56D03'}$, and there is no known persistent
X-ray source or globular cluster at this latitude.
Finally, the detection of an optical afterglow for this burst 
excludes an XRB interpretation.

The analysis of the two peaks separately does not give the
best fit. 
However, considering part (a) of the 1st peak,
we obtain a statistically significant improvement using the cutoff power-law
or the Band function, compared to the single power-law function, and the
spectrum has a significant break at $E_{\rm peak}$ (Table
\ref{tab:Spec_fit}).
 But in the other regions we cannot constrain the parameters by using the
Band function.

In the high energy band (above 10 keV), the photon index $\beta$ tends to be
less than -2, and smaller than the photon index $\alpha$.
$E^{\rm obs}_{\rm peak}$ can be constrained to be below 10 keV. But the band 
below 10 keV is near the low-energy threshold of WXM instrument
(2 keV).  If $E^{\rm obs}_{\rm peak}$ is near the low-energy threshold, 
the spectrum will appear to follow a power-law, even if it is actually
a Band function. 
We can constrain $E^{\rm obs}_{\rm peak}$ using a
$\textit{constrained}$ Band function. This can be done both for 
pure power-law spectra and power-law times exponential spectra with
the required curvature in the detector energy range, but only the
high-energy part of the Band function is allowed to produce a pure
power-law spectrum. This is described in Sakamoto et al
(2004).
This method is applicable when the spectra of the burst have
$E^{\rm obs}_{\rm peak}$ near or below the low-energy threshold of the
detector.

\begin{figure*}
  \begin{center}
    \FigureFile(80mm,60mm){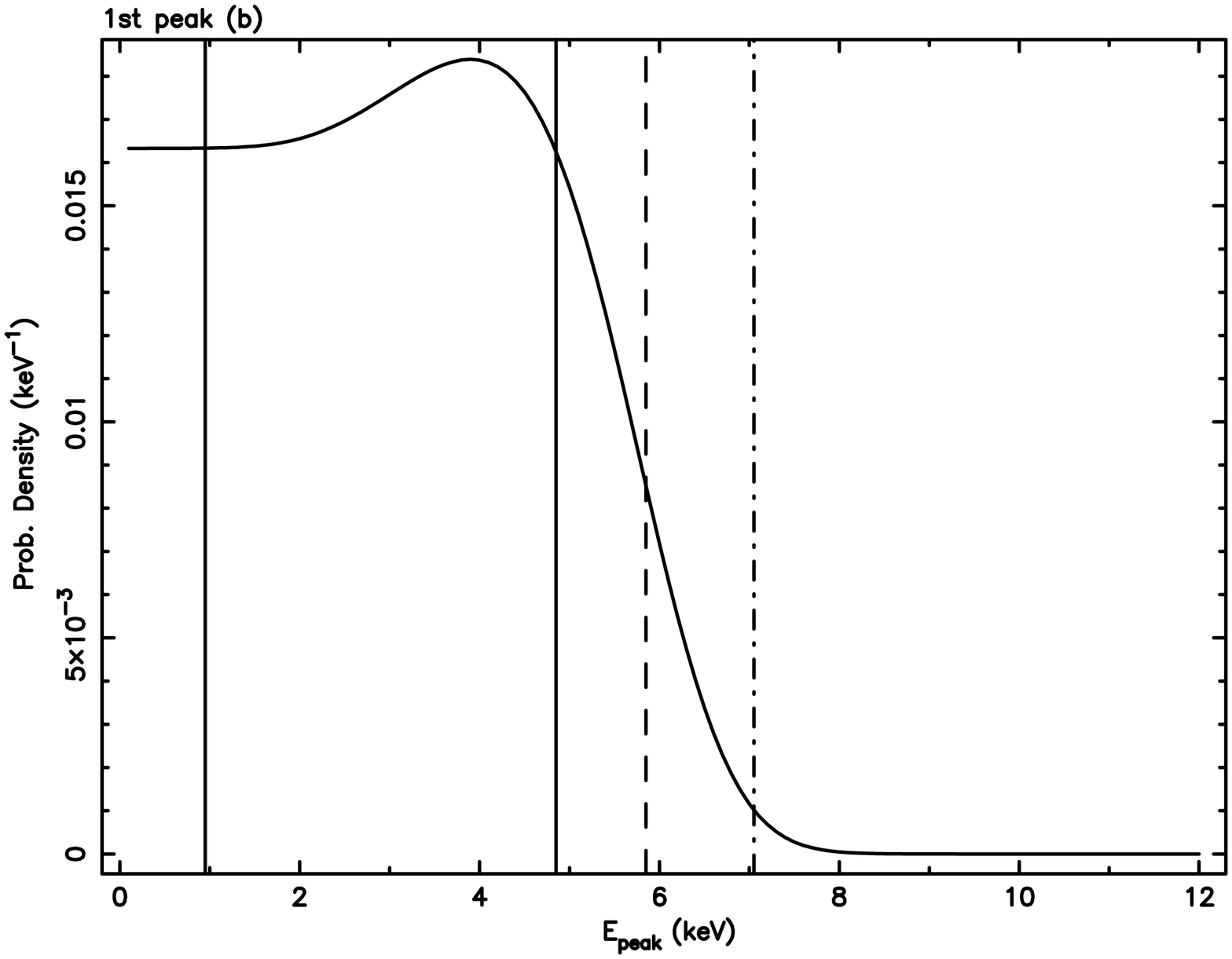}
    \FigureFile(80mm,60mm){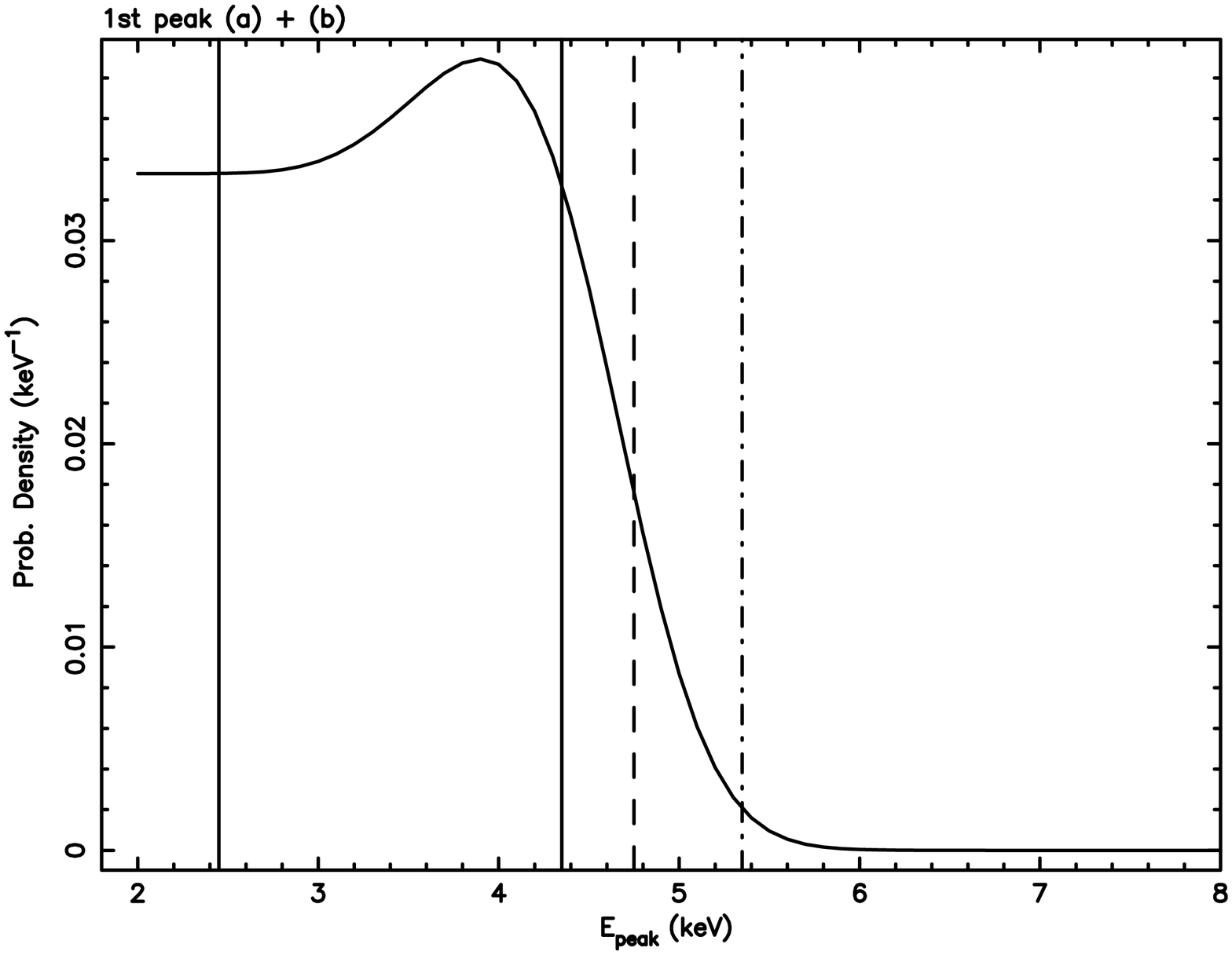}
    \FigureFile(80mm,60mm){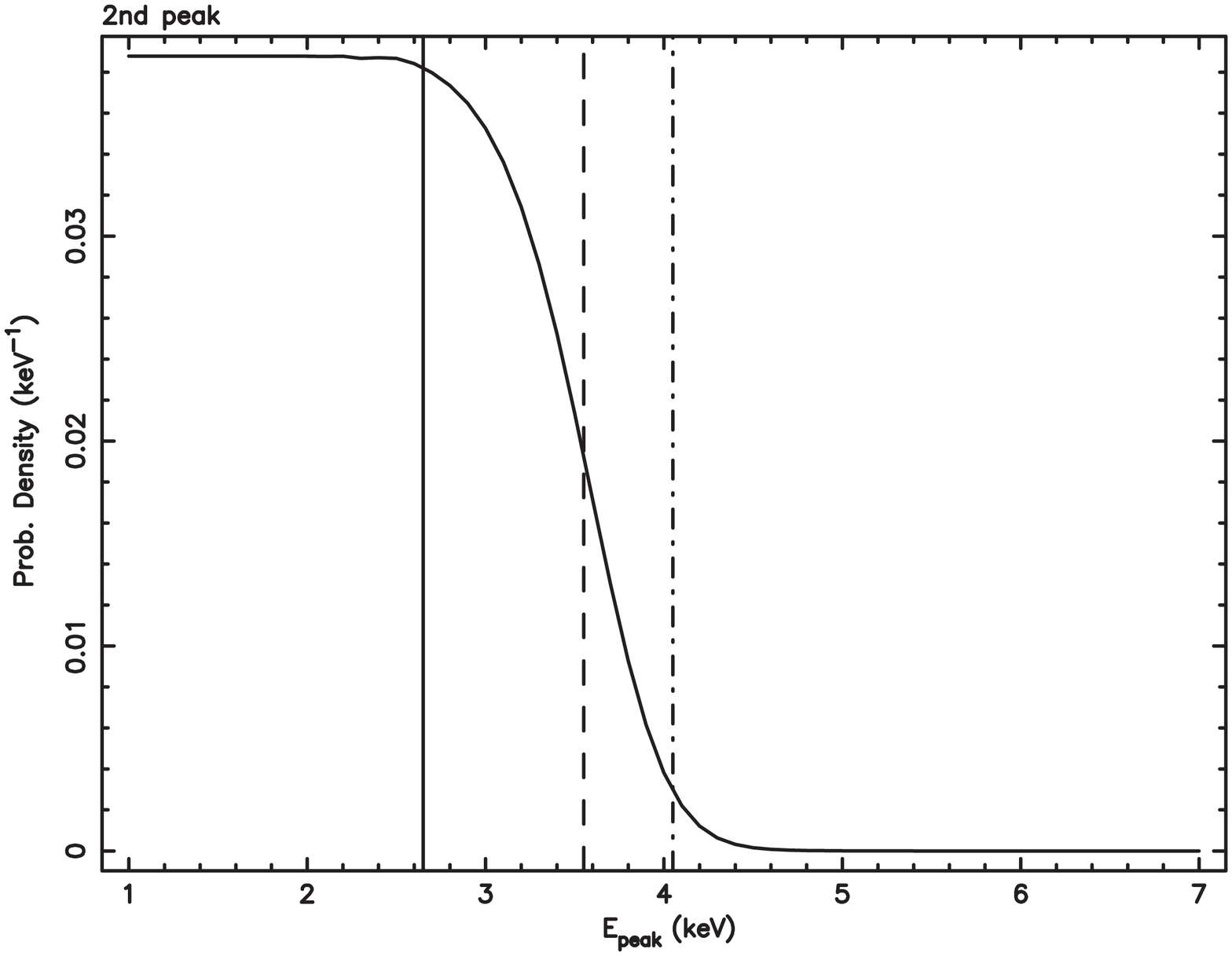}
    \FigureFile(80mm,60mm){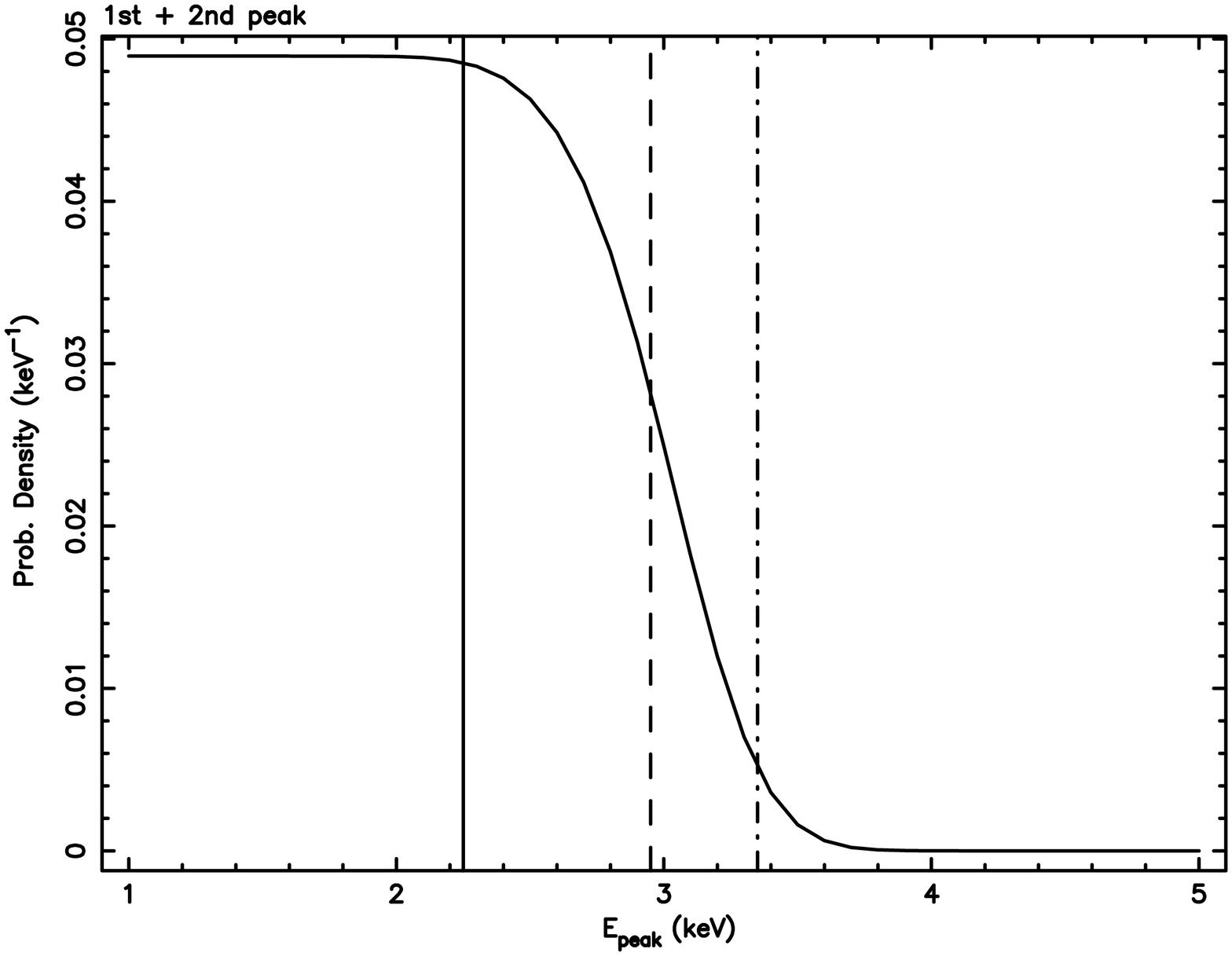}
  \end{center}
  \caption{Posterior probability density distribution for 
$E^{\rm obs}_{\rm peak}$. The vertical solid lines define the 68\%
 probatility interval for $E^{\rm obs}_{\rm peak}$, while the dashed and
 dotted lines show the 95\% and 99.7\% probability upper limits on 
 $E^{\rm obs}_{\rm peak}$.  Figures from top left to bottom right show
 the regions for the 1st peak (b), the 1st peak (a) + (b), the 2nd peak, and the 1st + 2nd peaks,
 respectively. } \label{fig:Constrained}
\end{figure*}

Applying the constrained Band function model to each
interval of XRF 040916, we obtained a constrained 
$E^{\rm obs}_{\rm peak}$.  The results are shown in Figure
\ref{fig:Constrained} as the
posterior probability density distribution for $E^{\rm obs}_{\rm peak}$.
From these distributions, we find best-fit
values for 68\%, 95\% and 99.7\% probabilities (Table \ref{tab:posterior}).
We conclude that XRF 040916 is extremely soft compared to typical
GRBs.  $E_{\rm peak}$ was determined for the 1st peak, but not 
for the 2nd (it is less than 4.8 keV
with 99.7\% probability); this hard-to-soft evolution is common in GRBs.

We also calculated the hardness ratio of the 30$-$400
keV ($S_{\gamma}$)  and 2$-$30 keV ($S_{\rm X}$)fluences. Using this
parameter, we can categorize bursts as
XRFs when $\log[S_{\rm X}/S_{\gamma}] > 0$,
X-ray rich GRBs when $-0.5 \leq \log[S_{\rm X}/S_{\gamma}] \leq 0$,
and classical hard spectrum GRBs when $\log[S_{\rm X}/S_{\gamma}] \leq -0.5$.
For the 1st peak, 
as $\log[S_{\rm X}/S_{\gamma}] \leq 0$ with 
the cutoff power-law function,
this burst would be classified as XRR rather than an XRF. 
For the 2nd peak, $\log[S_{\rm X}/S_{\gamma}]$ is 0.28, with a 90\%
confidence lower limit of 0.13, that is,
the emission of the 2nd peak is
softer than that of the 1st peak. 
Considering finally the total emission of the 1st and the 2nd peaks, 
the value of $\log[S_{\rm X}/S_{\gamma}]$, using the cutoff
power-law function, is 0.17 with a 90\% confidence lower limit of
0.06. Therefore we regard the entire burst as an XRF.

We consider the 1st peak to be clearly the prompt emission of XRF 040916;  we 
discuss the nature of the 2nd peak in the next section.

\begin{table*}
\begin{center}
 \caption{Results of the spectral analyses performed for XRF 040916}\label{tab:Spec_fit}
 (Note. The quoted errors correspond to 90\% confidence. )
 \begin{tabular}{lclclclclclclclcl}
  \hline\hline
  & & kT & & & $E^{\rm obs}_{\rm peak}$ & $\chi^2$ \\
  Region & Function &[keV] & $\alpha$ & $\beta$ & [keV] & (d.o.f.)  \\ \hline
%  \endfirsthead
  \hline
%  name & value \\
%  \endhead
 % \hline
%  \endfoot
 % \hline
 % \endlastfoot
  
  1st peak (a) ...... & blackbody & 5.4$^{+1.5}_{-1.2}$& & & & 21.68(21) & \\
  & Power law & & -1.6$^{+0.2}_{-0.2}$ & & & 24.41(21)  \\
  & Power law (above 10 keV)& & -2.0$^{+0.3}_{-0.5}$ & & & 8.76(13)  \\
  & Cutoff power law & & -0.39$^{+0.39}_{-0.95}$ & & 28.2$^{+48.2}_{-10.5}$ & 19.19(20)  \\
  & Band & & -0.16$^{+2.08}_{-1.03}$ & -2.4$^{+0.7}_{-7.6}$ & 25.6$^{+30.9}_{-12.5}$&
  18.00(19)  \\
  1st peak (b) ...... & blackbody & 1.2$^{+0.3}_{-0.2}$& & & & 25.57(21) \\
  & Power law & & -2.2$^{+0.4}_{-0.3}$ & & & 18.02(21)  \\
  & Power law (above 10 keV) & & -2.3$^{+0.7}_{-1.7}$ & & & 13.37(14) \\
  & Cutoff power law & & -1.9$^{+0.7}_{-...}$ $^a$ & &
  3.8$^{+3.7}_{-...}$ $^a$  & 18.37(20)  \\
  1st peak (a+b) ...... & blackbody & 2.3$^{+0.8}_{-0.6}$& & & & 55.60(34)  \\
  & Power law & & -1.8$^{+0.1}_{-0.2}$ & & & 26.00(34) \\
  & Power law (above 10 keV)& & -2.0$^{+0.3}_{-0.5}$ & & & 16.69(22)  \\
  & Cutoff power law & & -1.7$^{+0.3}_{-0.2}$ & & 79.3$^{+\infty}_{-59.5}$ & 25.66(33) \\
  2nd peak ...... & blackbody & 2.4  $^{+0.9}_{-0.8}$  & & & & 37.80(17)  \\
  & Power law & & -1.9$^{+0.2}_{-0.3}$ & & & 19.53(17) \\
  & Power law (above 10 keV)& & -2.5$^{+0.6}_{-1.3}$ & & & 10.13(10)  \\
  & Cutoff power law & & -1.6$^{+0.8}_{-0.5}$ & & 22.3$^{+\infty}_{-18.3}$ & 18.86(16) \\
  1st $+$ 2nd  ...... & blackbody & 1.8$^{+0.4}_{-0.4}$ & & & &
  82.30(47)  \\
  & Power law & & -1.9$^{+0.1}_{-0.2}$ & & & 42.16(47) \\
  & Power law (above 10 keV)& & -2.2$^{+0.3}_{-0.5}$ & & & 19.45(28) \\
  & Cutoff power law & & -1.8$^{+0.3}_{-0.3}$ & &
  25.8$^{+\infty}_{-21.5}$& 41.17(46) \\
  \hline
  \end{tabular}
\end{center}
 $^a$: The data do not allow a determination of the lower limits for
 the cutoff power-law function $\alpha$ or $E^{\rm obs}_{\rm peak}$ parameters.
\end{table*}

\begin{table*}
  \caption{Fluences using the cutoff power-law function}\label{tab:fluence}
  \begin{center}
    \begin{tabular}{c|c|c|c}
     \hline
      region &  $S_{\rm X}$ & $S_{\gamma}$ 
      & $\log[S_{\rm X}/S_{\gamma}]$ (90\% lower limit) \\ 
      &  [10$^{-7}$ erg cm$^{-2}$] & [10$^{-7}$ erg cm$^{-2}$] 
     & \\ \hline

     1st peak &  4.92$\pm$0.72 &  6.18$\pm$0.90 &   -0.10 ($>$-0.23) \\
     2nd peak &  2.91$\pm$0.50 &  1.53$\pm$0.26 &   0.28 ($>$0.13)\\
     1st peak + 2nd peak &  7.74$\pm$0.91 & 5.27$\pm$0.62 &  0.17 ($>$0.06)\\ \hline
    \end{tabular}
  \end{center}
\end{table*}

\begin{table*}
  \caption{Results of posterior probability density }\label{tab:posterior}
  \begin{center}
    \begin{tabular}{c|c|c|c|c}
     \hline
      & best-fit value (keV) & $E^{\rm obs}_{\rm peak}$ (keV) 
      & $E^{\rm obs}_{\rm peak}$ (keV)  & 
     $E^{\rm obs}_{\rm peak}$ (keV)  \\ 
      &  & with 68\%
     prob. &  with 95\% prob. & 
     with 99.7\% prob. \\ \hline

     1st peak (b) & 3.8 & 1.0 $-$ 4.9 & $<$5.9 & $<$7.1  \\
     1st peak (a) + (b) & 3.9 & 2.5 $-$ 4.4 & $<$4.8 & $<$5.5  \\
     2nd peak & ... & $<$2.7 &  $<$3.6 & $<$4.2  \\
     1st peak + 2nd peak & ... & $<$2.3 & $<$3.0 & $<$3.5  \\ \hline
    \end{tabular}
  \end{center}
\end{table*}

\section{Discussion}
\subsection{Redshift estimates}

The spectral softness could be explained if XRF 040916 were a high-redshift GRB. 
To check this hypothesis, we 
first estimated the redshift from the Amati relation (Amati et
al. 2002) using the 1st peak.
 Amati et al. (2002) found a relation between
the isotropic-equivalent radiated energy $E_{\rm iso}$ and the
burst-averaged value of $E_{\rm peak}$ in the rest frame
($E^{\rm rest}_{\rm peak}$ $\propto$ $E_{\rm iso}^{0.5}$, with 
$E^{\rm rest}_{\rm peak}$
in keV and $E_{\rm iso}$ in units of $10^{52}$ ergs). 

\begin{figure*}
  \begin{center}
    \FigureFile(88mm,45mm){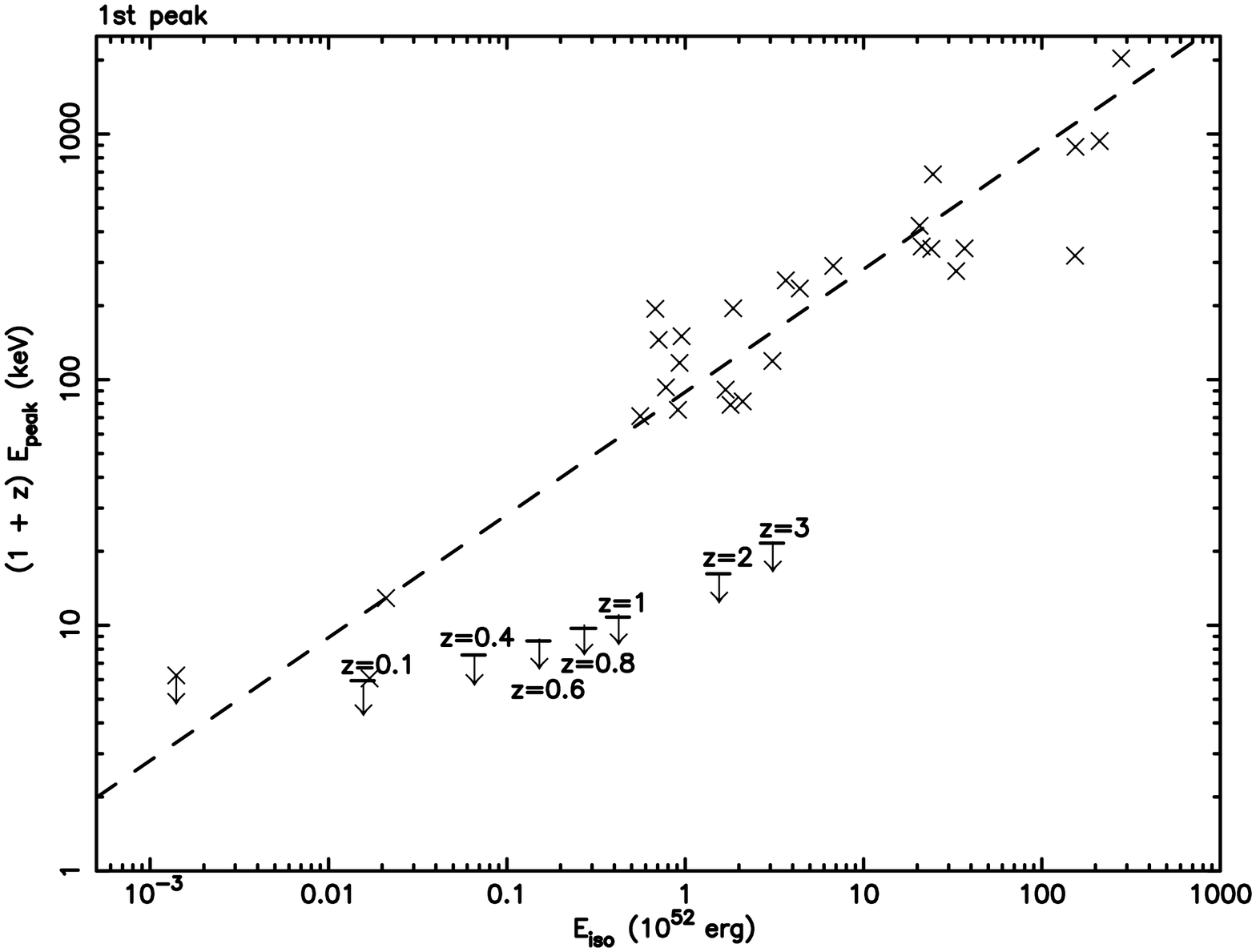}
    \FigureFile(88mm,45mm){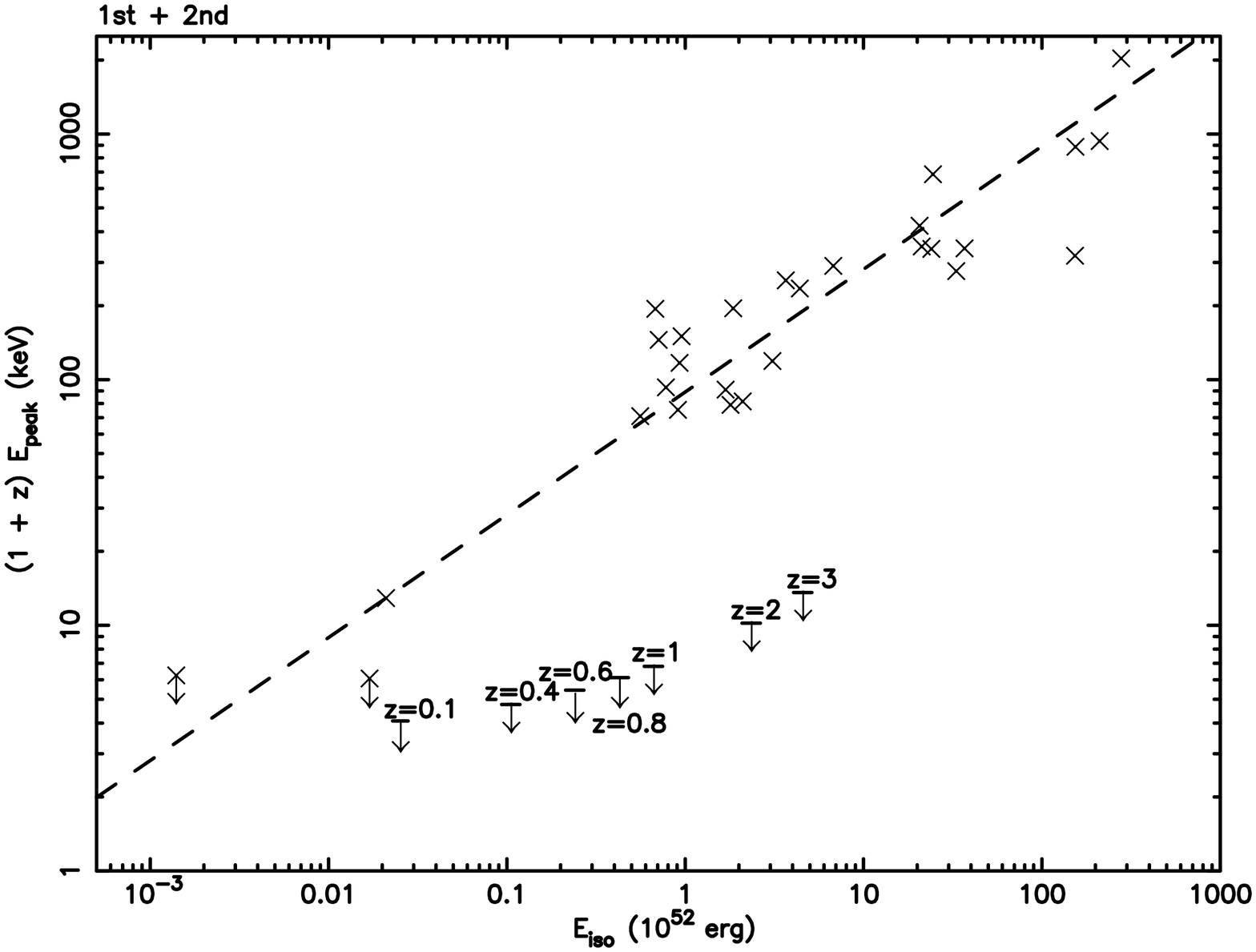}
    %%% \FigureFile(width,height){filename}
  \end{center}
  \caption{The ($E_{\rm iso}$-$E^{\rm rest}_{\rm peak}$) -plane where
 $E_{\rm iso}$ is the isotropic-equivalent radiated energy and 
 $E^{\rm rest}_{\rm peak}$ is the peak energy of the $\nu F_{\nu}$ spectrum,
 both measured in the
 rest frame of the burst. The bars are for XRF 040916 at
 various distances. The crosses are the $\textit{BeppoSAX}$-GRBs (Amati et
 al. 2002) and $\textit{HETE-2}$ GRBs. The dashed line is the equation,
 $E^{\rm rest}_{\rm peak}$ = 89($E_{\rm iso}$/10$^{52}$ ergs)$^{0.5}$
 keV given by Amati et al. (2002).}
  \label{Fig:Amati}
\end{figure*}

Assuming this relation, it is possible to estimate 
the  redshift using  only the flux and
$E_{\rm peak}$ of a burst.
We calculated $E^{\rm rest}_{\rm peak}$ and $E_{\rm iso}$
from the spectral parameters of the 1st peak assuming
various redshifts.
As shown in Figure \ref{Fig:Amati},
the smaller the redshift is, the more consistent the
computed values are with the Amati relation.
This result is also consistent with the Subaru redshift constraint of $z < 3$,
imposed by its detection of the  optical afterglow in the B-band(Kosugi et
al. 2004)

We have also computed an upper limit to
the pseudo-redshift (with the method described in P{\'e}langeon et al. 2006)
using the WXM spectrum for the  most intense 15s long part
of the 1st peak.
To do this, we derived the upper limit to $E_{\rm peak}$ using the constrained
Band model to fit the data.
We find that $E_{\rm peak}$ $<$ 6.2 keV with 90\% confidence, 
leading to a pseudo-redshift $<$ 0.7, consistent with the
preceding results.  Thus we can reject a high redshift for
this burst.

\subsection{Origin of the 2nd peak}
 
In the spectral analysis, we could not determine $E_{\rm peak}$ for the 2nd
peak, but we found that above 10 keV the best fitting
model was a power-law, and  that $E_{\rm peak}$ lay in the low energy region
($<$4keV). Furthermore, the 1st and 2nd peaks are separated by 
$\sim$ 200 s.
This suggests that the 
2nd peak 
could be the X-ray afterglow rather than prompt emission.
Indeed there are at least five possibilities to explain it. 
 (1) beginning of the afterglow
 (2) Ambient density
fluctuations of the X-ray afterglow, 
 (3) Patchy shell, 
 (4) Refreshed shock and (5) Long-acting engine model
(Ioka et al. 2005).  We consider them in the
following sections.
\\

\subsubsection{Beginning of the afterglow}
Piro et al. (2005) said that GRB 011121 and GRB 011211 showed a
late X-ray burst taking place a few hundreds of seconds after the prompt
emission. The spectral and temporal evolution of the afterglow indicated
that fireball evolution took place in the interstellar medium for GRB
011211 and in the wind for GRB 011121 respectively.
In both cases, the decay time of a late X-ray burst (i.e. afterglow) is 
supposed to be proportional to $\sim t^{-1}$ (Sari et al. 1998). 

If we set the zero time ($t_0$) of the emission episode
for the 1st peak, then 
the decay time of the 2nd peak is proportional to $t^{-12.7\pm0.3}$
in 2$-$25 keV energy band. 
Since this result is inconsistent with the above
afterglow model, we can reject the possibility of the beginning of the
afterglow.
\\

\subsubsection{Ambient density fluctuation model}
This model explains
afterglow variabilities by ambient density fluctuations caused by
turbulence in the interstellar medium or variable winds from the
progenitor star. 
If XRF 040916 was in the afterglow stage
when the 2nd peak occurred, we can use the following formula for a
kinematical upper limit on the variability  (Ioka et al. 2005)
\begin{equation}
\frac{\Delta F_\nu}{F_\nu} \le 2 f_c^{-1}
 \frac{F}{\nu F_\nu} \biggl(\frac{\Delta t}{t}\biggr)^2, \label{eq:Ioka}
\end{equation}
where $\Delta t$ is the variability timescale, $t$ is the observed time
since the burst, $F_\nu$ is the assumed power-law baseline of the
afterglow flux, $F$ is the bolometric base flux, $f_c$ $\sim$
($\nu_m$/$\nu_c$)$^{(p-2)/2}$ where the cooling frequency is $\nu_c$, the
characteristic synchrotron frequency is $\nu_m$ , and the electron power-law
distribution index is $p$,  $F/ \nu F_\nu$ $\sim$
($\nu$/$\nu_c$)$^{(p-2)/2}$ , and
 $\Delta t$, $\Delta F_\nu$ are the timescale and amplitude
deviations above the baseline, respectively.
This formula gives the maximum possible afterglow variability due to
ambient density fluctuations.
We can estimate the factor $F/ \nu F_\nu$ assuming the standard
afterglow model for $\nu_m < \nu_c < \nu$ (the X-ray band at $t \sim$ 370s).
Since $\nu_c$ $\sim$ 10$^{16}$ Hz at t $\sim$ 370s for $p$ $\sim$ 2.2
(Sari et al. 1998), we have $F/ \nu F_\nu$ $\sim$ 1 for the X-ray band
($\nu$ $\sim$ 10$^{19}$ Hz). 
Furthermore, substituting $\Delta t \sim$ 60 s in equation
\ref{eq:Ioka},
 the right-hand side becomes $\sim$ 0.1.
But we can clearly see that 
$\Delta F_\nu / F_\nu$ $\gg$ 1 from the flux variabilities in 
Figure \ref{fig:thx}, because the flux $F\nu$ in the region between
the 1st and 2nd peaks ($<$ 1.8 $\times$ 10$^{-9}$ ergs cm$^{-2}$
s$^{-1}$ with 2$\sigma$ confidence level,
2$-$400 keV) which
is assumed to be the afterglow stage, is less than the flux of the 2nd peak (8.3
$\times$ 10$^{-9}$ ergs cm$^{-2}$ s$^{-1}$, 2$-$400 keV) 
by a factor of more than  5 with 2$\sigma$ confidence level.
Since the observed variability exceeds the maximum allowed value,
ambient density fluctuations
cannot explain the nature of the 2nd peak.
\\

\subsubsection{Patchy shell model}
In the patchy shell
model the variability timescale of the afterglow at time $t$ must 
be $\Delta t$ $\ge$ $t$ (Nakar \& Oren 2004). 
In this model, the GRB jet consists of
many subjets (Yamazaki et al. 2004;
Ioka \& Nakamura 2001). Since we observe an angular size $\gamma^{-1}$ within a
GRB jet with Lorentz factor $\gamma$, the flux depends on the
angular structure for the observer.
The patchy shell model cannot make a bump with variability timescale 
$\Delta t \sim $ 60 s, which is shorter than the observed timescale $t \sim $ 370
s, so
we can marginally reject this model from the point of view of the
variability timescale.
In addition, the flux of the 2nd peak is as bright as that of the 1st one.
If this burst is explained by the patchy shell model, we must assume
very large shell non-uniformity.
In this case, we have to assume that we first observe the dark part of
the shell,
and then the bright part of the shell as the Lorentz factor drops.
As  the assumption of a large non-uniformity shell is unrealistic,
 we can consequently exclude
the patchy model.
\\

\subsubsection{Refreshed shock model}
In this model, multiple shells are ejected at various velocities, and
the variability occurs when the slow inner shell catches up with the fast
outer shell a long time later, since the velocity of the outer shell
decreases through the interaction with the ambient medium (Rees \&
M\'esz\'aros 1998; Panaitescu et
al. 1998; Kumar \& Piran 2000; Sari \& M\'esz\'aros 2000; Zhang \&
M\'esz\'aros 2002). The variability timescale at time $t$  is given by
$\Delta$ $t$ $\ge$ $t$/4 (Ioka et al. 2005). In the case of XRF 040916,
however, the variability timescale is $\Delta$ $t$ $\le$ $t$/4 for $t$
$\sim$ 370 s and $\Delta t$ $\sim$ 60 s, so we can marginally reject the
refreshed shock model.

In addition, the flux variability will not be equal to the GRB flux.
The relation between the energy increase factor $F$ and the flux increase
$f$ is given by $f = F^{(3+p)/4}$ for $\nu_{m} < \nu < \nu_c$, and
$f = F^{(2+p)/4}$ for $\nu_{m}, \nu_{c} < \nu $ (Granot et al. 2003). 
In either case, 
$f \sim F$ with typical values of $p \sim$ 2.2.  Because 
$f \gg 1$ from the observation, we must have $F \gg 1$. This means that the
slow shells have very large energy, which contradicts the refreshed
shock model (Refreshed shocks produce changes of less than 1 order of
magnitude). Therefore this model, too, can be rejected.
\\

\subsubsection{Long-acting engine model}
In this
model, at the observed time $t$, the central engine is still active and emitting
shells (Rees \& Meszaros 2000; Zhang \& Meszaros 2002; Dai \&
Lu 1998). This can explain variability timescales
down to a millisecond and there is no restriction
on the flux variability. The most likely explanation of the 2nd peak is
therefore the long-acting engine model. Only this model can explain both
the variability timescale and the flux variability for the 2nd peak.
In this scenario, both the 1st and 2nd peaks have the same mechanism,
and both must show the same spectral and temporal features as GRBs.
As we have shown in Section \ref{sec:temporal}, the timescale of
the 2nd peak in Table \ref{tab:temporal_tot} is shorter at higher
energies, which is a typical feature of GRBs.

Furthermore, we considered the curvature effect (Kumar \& Panaitescu
2000; Liang et al. 2006). This effect is the rapid decay due to the
observed receiving the progressively delayed emission from higher
latitudes. The following formula represents the curvature effect,
\begin{equation}
F_\nu(t) = A\biggl(\frac{t-t_0}{t_0}\biggr)^{-(1+\beta)} , \label{eq:Curvature}
\end{equation}
where $\beta$ is the X-ray photon index during the decay, $t_0$ is the
time zero point of the emission episode related to decay and $A$ is 
normalization parameter for decay component.
We applied this effect to the 2nd peak. Because of the poor statistics 
in the 2nd peak, we used the X-ray photon index ($\beta$ = 1.9) during the
entire 2nd peak (rise and decay) with the background subtracted in the
region (2).
From fitting analysis, we obtained $t_0$ = -34.8 $\pm$ 34.9 s,
near the time zero point for the 2nd peak.
This result implies that the zero time for the 1st peak doesn't coincide with
the 2nd peak one: then the central engine was reactivated.

And as we showed in Section \ref{sec:spectrum}, the
spectrum of the 2nd peak is probably softer than that of the 1st. Thus
$E_{\rm peak}$ appears to decrease with time and flux, in accord with
the well known
hard-to-soft evolution in GRB spectra (Fishman \& Meegan 1995). This
feature is also consistent with the long-acting engine model, and we consider
it to be the most reasonable mechanism to explain
the soft 2nd peak of XRF 040916.

Furthermore, we can compare this
event with the Swift bursts XRF 050406 and GRB 050502B.
Swift (Gehrels et al. 2004) can
detect early X-ray afterglows with the X-Ray Telescope (XRT;
Burrows et al. 2005a) and it has detected 
bright X-ray flares 
100$-$1000 seconds after the prompt emission for these bursts.  They have been 
explained by the long-acting engine model (Burrows et al. 2005b, Romano et
al. 2006, Falcone et al. 2006).   
Because the XRT is only sensitive to photons in the 0.2$-$10 keV energy range,
the spectral features of the X-ray flares are not as well resolved
as those of the prompt emission, and their emission mechanism is still under discussion. 
In GRB 011211( Jakobsson et al. 2004; Holland
et al. 2002;  Piro et al. 2005),  GRB 011121 (Piro et
al. 2005) and GRB 021004( Bersier et al. 2003; Halpern et
al. 2002) the afterglows included X-ray flares, whose study
revealed a wealth of information about the central engine and its
surrounding regions.
For XRF 050406, the bright flare occurred $\sim$200 s
after the prompt emission, and for GRB 050502B, it occurred $\sim$700
s later . The hard to soft count rate ratios for the 
flares was similar to the spectral evolution of the prompt GRB emission.
As the time interval of XRF 040916 is also
of the order of $\sim$200 s and the emission of the 1st peak is harder
than that of the 2nd according to the $\log[S_{\rm X}/S_{\gamma}]$ ratio,
the timescale and the spectral evolution between the peaks for XRF
040916 is similar to
those of XRF 050406 and GRB 050502B, and can be explained by
the long-acting engine model.

\section{Conclusion}
In this paper we have reported the \textit{HETE-2} WXM/FREGATE
observations of the
XRF 040916, which consists of two peaks  separated by $\sim$200 s. 
For this burst, only an upper limit to the redshift has been reported
based on optical observations ($ z < 3$). Taking into account this
limit,
 our estimate from the Amati relation
is consistent with a small redshift.
In terms of the spectral evolution, it seems that the 2nd peak is softer than the
1st peak. 
We have studied the emission mechanism of the 2nd
peak and discussed its most probable origin.
Considering different models 
( beginning of the afterglow, ambient density fluctuations, patchy shell, and refreshed shock model),
we found that the long-acting engine model is the most plausible one to
explain both the timescale and flux variabilities.
In some GRBs or XRFs
(e.g. GRB050502B and XRF 050406) (Burrows et al. 2005b), bright X-ray
flares were observed and these bursts are also explained by the
long-acting engine model. 
Consequently, our results indicate that the case of XRF 040916 
is similar to those of the X-ray flares detected by Swift.  
\\

We are grateful to K. Ioka for giving us a fruitful advice to develop the
discussion.
We greatly appreciate the anonymous referee for his/her comments and
suggestions that improved this paper. We would like to thank the
$\textit{HETE-2}$ team members for their support. The \textit{HETE-2} mission is
supported in the US
by NASA contract NASW-4690; in Japan in part by Grant-in-Aid 14079102
from the Ministry of Education, Culture, Sports, Science, and
Technology; and in France by CNES contract 793-01-8479.
KH is grateful for support under MIT contract MIT-SC-R-293291.
This work was supported by a 21st Century COE Program at TokyoTech
"Nanometer-Scale Quantum Physics" by the Ministry of Education, Culture,
Sports, Science and Technology.

%%%%%%%%%%%%%%%%%%%%%%%%%%%%%%%%%%%%%%%

%\appendix
%\section{Method of .....}

%\section{Approximation of ...}

%\section*{Complete data}

%%%
% See the manual for the detail.
%%%


\begin{thebibliography}{}
% Journals(e.g. A\&A,ApJ,AJ,NMRAS,PASP ...)
\bibitem[Arnaud(1996)]{1996ASPC..101...17A} Arnaud, K.~A.\ 1996, ASP 
Conf.~Ser.~101: Astronomical Data Analysis Software and Systems V, 101, 17 
\bibitem[Amati et al.(2002)]{2002A&A...390...81A} Amati, L., et al.\ 2002, 
\aap, 390, 81 
\bibitem[Atteia et al.(2003)]{2003AIPC..662...17A} Atteia, J.-L., et al.\ 
2003, AIP Conf.~Proc.~662: Gamma-Ray Burst and Afterglow Astronomy 2001: A 
Workshop Celebrating the First Year of the HETE Mission, 662, 17 
\bibitem[Band et al.(1993)]{1993ApJ...413..281B} Band, D., et al.\ 1993, 
\apj, 413, 281 
\bibitem[Barraud et al.(2003)]{2003A&A...400.1021B} Barraud, C., et al.\ 
2003, \aap, 400, 1021
\bibitem[Bersier et al.(2003)]{2003ApJ...584L..43B} Bersier, D., et al.\ 
2003, \apjl, 584, L43 

\bibitem[Burrows et al.(2005)]{2005SSRv..120..165B} Burrows, D.~N., et al.\ 
2005a, Space Science Reviews, 120, 165 
\bibitem[Burrows et al.(2005)]{2005Sci...309.1833B} Burrows, D.~N., eqt al.\ 
2005b, Science, 309, 1833 
\bibitem[Dai \& Lu(1998)]{1998A&A...333L..87D} Dai, Z.~G., \& Lu, T.\ 1998, 
\aap, 333, L87 
\bibitem[Falcone et al.(2006)]{2006ApJ...641.1010F} Falcone, A.~D., et al.\ 
2006, \apj, 641, 1010 
\bibitem[Fenimore et al.(1995)]{1995ApJ...448L.101F} Fenimore, E.~E., in 't 
Zand, J.~J.~M., Norris, J.~P., Bonnell, J.~T., \& Nemiroff, R.~J.\ 1995, 
\apjl, 448, L101 
\bibitem[Fishman \& Meegan(1995)]{1995ARA&A..33..415F} Fishman, G.~J., \& 
Meegan, C.~A.\ 1995, \araa, 33, 415 
\bibitem[Gehrels et al.(2004)]{2004ApJ...611.1005G} Gehrels, N., et al.\ 
2004, \apj, 611, 1005 
\bibitem[Granot et al.(2003)]{2003Natur.426..138G} Granot, J., Nakar, E., 
\& Piran, T.\ 2003, \nat, 426, 138 
\bibitem[Halpern et al.(2002)]{2002GCN..1578....1H} Halpern, J.~P., 
Armstrong, E.~K., Espaillat, C.~C., \& Kemp, J.\ 2002, GRB Coordinates 
Network Circ., 1578, 1 
\bibitem[Henden(2004)]{2004GCN..2722....1H} Henden, A.\ 2004a, GRB 
Coordinates Network Circ., 2722, 1 
\bibitem[Henden(2004)]{2004GCN..2727....1H} Henden, A.\ 2004b, GRB 
Coordinates Network Circ., 2727, 1 
\bibitem[Holland et al.(2002)]{2002AJ....124..639H} Holland, S.~T., et al.\ 
2002, \aj, 124, 639 
\bibitem[Ioka \& Nakamura(2001)]{2001ApJ...554L.163I} Ioka, K., \& 
Nakamura, T.\ 2001, \apjl, 554, L163 
\bibitem[Ioka et al.(2005)]{2005ApJ...631..429I} Ioka, K., Kobayashi, S., 
\& Zhang, B.\ 2005, \apj, 631, 429 
\bibitem[Jakobsson et al.(2004)]{2004NewA....9..435J} Jakobsson, P., et 
al.\ 2004, New Astronomy, 9, 435 
\bibitem[Kosugi et al.(2004)]{2004GCN..2726....1K} Kosugi, G., Kawai, N., 
Tajitsu, A., \& Furusawa, H.\ 2004a, GRB Coordinates Network Circ., 2726, 1 
\bibitem[Kosugi et al.(2004)]{2004GCN..2730....1K} Kosugi, G., Kawai, N., 
Tajitsu, A., \& Furusawa, H.\ 2004b, GRB Coordinates Network Circ., 2730, 1 
\bibitem[Kumar \& Panaitescu(2000)]{2000ApJ...541L..51K} Kumar, P., \& 
Panaitescu, A.\ 2000, \apjl, 541, L51 
\bibitem[Kumar \& Piran(2000)]{2000ApJ...532..286K} Kumar, P., \& Piran, 
T.\ 2000, \apj, 532, 286 
\bibitem[Liang et al.(2006)]{2006ApJ...646..351L} Liang, E.~W., et al.\ 
2006, \apj, 646, 351 
\bibitem[Nakar \& Oren(2004)]{2004ApJ...602L..97N} Nakar, E., \& Oren, Y.\ 
2004, \apjl, 602, L97 
\bibitem[Panaitescu et al.(1998)]{1998ApJ...503..314P} Panaitescu, A., 
Meszaros, P., \& Rees, M.~J.\ 1998, \apj, 503, 314 
\bibitem[P{\'e}langeon \& The Hete-2 Science 
Team(2006)]{2006AIPC..836..149P} P{\'e}langeon, A., \& The Hete-2 Science 
Team 2006, American Institute of Physics Conference Series, 836, 149 
\bibitem[Piro et al.(2005)]{2005ApJ...623..314P} Piro, L., et al.\ 2005, 
\apj, 623, 314 
\bibitem[Rees \& M{\'e}sz{\'a}ros(1998)]{1998ApJ...496L...1R} Rees, M.~J., \& 
M{\'e}sz{\'a}ros, P.\ 1998, \apjl, 496, L1 
\bibitem[Rees \& M{\'e}sz{\'a}ros(2000)]{2000ApJ...545L..73R} Rees, M.~J., 
\& M{\'e}sz{\'a}ros, P.\ 2000, \apjl, 545, L73 
\bibitem[Romano et al.(2006)]{2006A&A...450...59R} Romano, P., et al.\ 
2006, \aap, 450, 59 
\bibitem[Sakamoto et al.(2004)]{2004ApJ...602..875S} Sakamoto, T., et al.\ 
2004, \apj, 602, 875 
\bibitem[Sakamoto et al.(2005)]{2005ApJ...629..311S} Sakamoto, T., et al.\ 
2005, \apj, 629, 311 
\bibitem[Sari et al.(1998)]{1998ApJ...497L..17S} Sari, R., Piran, T., \& 
Narayan, R.\ 1998, \apjl, 497, L17 
\bibitem[Sari \& M{\'e}sz{\'a}ros(2000)]{2000ApJ...535L..33S} Sari, R., \& 
M{\'e}sz{\'a}ros, P.\ 2000, \apjl, 535, L33 
\bibitem[Shirasaki et al.(2003)]{2003PASJ...55.1033S} Shirasaki, Y., et 
al.\ 2003, \pasj, 55, 1033 
\bibitem[Villasenor et al.(2003)]{2003AIPC..662...33V} Villasenor, J.~N., 
et al.\ 2003, AIP Conf.~Proc.~662: Gamma-Ray Burst and Afterglow Astronomy 
2001: A Workshop Celebrating the First Year of the HETE Mission, 662, 33 
\bibitem[Yamamoto et al.(2004)]{2004GCN..2713....1Y} Yamamoto, Y., et al.\ 
2004, GRB Coordinates Network Circ., 2713, 1 
\bibitem[Yamazaki et al.(2004)]{2004ApJ...607L.103Y} Yamazaki, R., Ioka, 
K., \& Nakamura, T.\ 2004, \apjl, 607, L103 
\bibitem[Yamazaki et al.(2004)]{2004GCN..2712....1Y} Yamazaki, T., et al.\ 
2004, GRB Coordinates Network Circ., 2712, 1 
\bibitem[Zhang \& M{\'e}sz{\'a}ros(2002)]{2002ApJ...566..712Z} Zhang, B., 
\& M{\'e}sz{\'a}ros, P.\ 2002, \apj, 566, 712 




% Books
%\bibitem[Author(2001)]{key-2}
%   Author A.A., Author B.B.\ 2001, Name of Book(Publisher, Tokyo) ch0
% Books
%\bibitem[Author(2001)]{key-3}
%  Author A.A., Author B.B.\ 2001, Name of Book(Publisher, Tokyo) page0

% Editorial Books
%\bibitem[Author(2001)]{key-n}
%  Author A.A.\ 2001, in Name of Book,
%   ed Editor D.\ Editor(Publisher, Tokyo) page0
\end{thebibliography}
\end{document}